\documentclass[10pt]{iopart}
\usepackage{iopams}  
\usepackage{graphicx} 
\begin{document}

\title{Production of highly charged ions inside a cryogenic Penning trap 
by electron-impact ionisation}

\author{Kanika$^{1,2*}$, A Krishnan$^{1,3*}$, J W Klimes$^{1,2,4}$, B Reich$^{1,2}$, K K Anjum$^{1,5}$, P Baus$^3$, G Birkl$^{3,6}$,  W Quint$^{1,2}$  and M Vogel$^1$}

\address{$^1$ GSI Helmholtzzentrum für Schwerionenforschung, Planckstrasse 1, 64291 Darmstadt, Germany}
\address{$^2$ Universität Heidelberg, Grabengasse 1, 69117 Heidelberg, Germany}
\address{$^3$ Technische Universit\"at Darmstadt, Institut f\"ur Angewandte Physik, Schlossgartenstra\ss{}e 7, 64289 Darmstadt, Germany}
\address{$^4$ Max-Planck-Institut für Kernphysik Heidelberg, Saupfercheckweg 1, 69117 Heidelberg, Germany}
\address{$^5$ Friedrich-Schiller-Universit\"at Jena, Max-Wien-Platz 1, 07743 Jena, Germany} 
\address{$^6$ Helmholtz Forschungsakademie Hessen f\"ur FAIR (HFHF), Campus Darmstadt, Schlossgartenstra\ss e 2, 64289 Darmstadt, Germany}

\ead{k.kanika@gsi.de}
\ead{a.krishnan@gsi.de}

\vspace{10pt}
\begin{indented}
\item[]*Shared first authorship 
\\
\item[]April 2023

\end{indented}

\begin{abstract}
We have built and operated a cryogenic Penning trap arrangement that allows for the efficient production, selection, and long-term storage of highly charged atomic ions. In close similarity to an electron-beam ion trap (EBIT) it works by electron-impact ionisation of atoms inside a dedicated confinement region. The electrons are produced by field emission at liquid-helium temperature and are subsequently accelerated to the keV energy range. The electron beam is reflected through the trap multiple times to increase the ionisation efficiency. We show a characterisation of the system and measurements with argon and tungsten ions up to Ar$^{16+}$ and W$^{27+}$, respectively.
\end{abstract}

\vspace{2pc}

\ioptwocol
\section{Introduction}
Highly charged atomic ions are sought-after study objects in several contexts \cite{null}, particularly with regard to precision measurements of X-ray, optical and microwave transitions in few-electron systems that are bound by the extreme electromagnetic fields of the atomic nucleus \cite{3}. This area of research comprises measurements of fine- and hyperfine-structure energies and lifetimes \cite{micke}, of the Lamb shift \cite{gumb,dumb} and the bound-electron magnetic moments \cite{1,2,2b} as benchmarks of bound-state quantum electrodynamics \cite{3,3b,3c}, as well as spectroscopy in the framework of metrology \cite{4} and the determination of fundamental constants \cite{5,6}. Much of this effort involves charged-particle traps such as Penning traps \cite{19} for long-term confinement and preparation of the desired species under well-defined conditions, usually in cryogenic surroundings and at extreme vacua. 

Depending on the charge state under consideration, the highly charged ions may be produced by various methods, most of which rely on an external source, combined with subsequent transport and capture into a Penning trap \cite{7,8,9,9b}. In contrast to singly charged ions, medium and high charge states cannot be efficiently produced by photo-ionisation inside the trap, as the required intensities are far beyond readily available lasers. However, for charge states up to ionisation potentials of a few keV, it is possible to create the ions inside a separate potential well of a Penning trap that is operated in similarity to an electron-beam ion trap (EBIT): electrons from a field-emission point or -array \cite{array} are accelerated such that they traverse a dedicated trap well and ionise atoms that are present there \cite{10}. In similarity to an EBIT operated in reflex mode \cite{donets}, the electron beam in the present setup is reflected through the trap multiple times, thereby increasing the production efficiency. The ion content of this trap can be monitored non-destructively until the targeted charge-state distribution is reached. Ions can then be selected by their charge-to-mass ratio and the resulting pure ensemble can be moved to an adjacent trap for the intended measurements. 

In the present case, this is laser-microwave double-resonance spectroscopy of a large and cooled single-species ensemble of highly charged ions that aims to measure the magnetic moments ($g$-factors) of bound electrons and atomic nuclei with high precision \cite{2b,art2}. Such measurements represent valuable benchmarks of calculations in the framework of the quantum electrodynamics of bound states, i.e. in the presence of extreme electromagnetic fields \cite{1,2}. This work is part of the ARTEMIS experiment \cite{2b,art2} located at the HITRAP facility \cite{hitrap,hit2} at GSI, Germany. The present measurements use the existing ARTEMIS Penning trap equipped with the internal ion source under discussion \cite{dvl} as it has been used previously for measurements of highly-charged-ion cooling \cite{ebra}. In the present article, we show the setup and procedures of in-trap ion production, and present characterising measurements with ions up to Ar$^{16+}$ and W$^{27+}$. Together with electron-impact ionisation simulations, these are used to make quantitative statements about the electron beam and the performance of the device, particularly with regard to the operation in reflex mode.  

\section{Background: Theory and Methods}
\subsection{Field Emission of Electrons from a Cryogenic Field-Emission Point (FEP)}
In a small cryogenic environment, a favourable source of electrons for ionisation of atomic particles is a field-emission point or -array \cite{array}, since it is compact and not connected with a thermal load on the environment. A voltage of typically a few kV is applied to a metallic needle (usually made of tungsten) with a tip of size of the order of 100\,nm or less. The corresponding electric field from the tip to the earthed surroundings enhances quantum tunneling of electrons into the vacuum which are hence emitted from the tip surface \cite{11}. This is commonly called a `Müller-type' field emitter \cite{12}. For a given needle, the achieved electron current can in principle be determined from the tip geometry and the applied voltage by Fowler-Nordheim-type equations \cite{11}, but for real emitters it is more reliable to obtain this number from a measurement as will be discussed below. Typically achievable electron currents from a single cryogenic FEP are of the order of $\mu$A and below.

\subsection{Electron-Impact ionisation in EBITs}
An EBIT uses electrons to create highly charged ions by subsequent electron-impact ionisation (`charge breeding') of a given atomic system, either an atom or an ion of a lower charge state, that is confined by suitable electromagnetic fields \cite{13}. The electron kinetic energy must exceed the ionisation potential (IP) of the system for its next charge state to be produced. For optimum production, it is commonly chosen around the maximum of the electron-impact ionisation cross section which is usually between 2 and 3 times the IP \cite{13b}. During production, for any present charge state there is gain by ionisation from the lower charge states, loss by ionisation to higher charge states, and loss due to several processes that result in electron capture. Overall, the situation can be quantified by a set of coupled rate equations, one for each charge state present \cite{14}. Each equation has gain and loss terms, a detailed description requires knowledge (for each individual charge state) about the ionisation cross section at the given electron energy and about the cross sections for the loss mechanisms: radiative recombination \cite{15}, charge exchange \cite{16}, and the loss of confinement. With these quantities known, the temporal evolution of a given charge state distribution under the influence of an electron beam at a given kinetic energy $E$ and current density $j$ can be obtained from corresponding computer codes \cite{17,18}. 

\subsection{Ion Confinement in a Penning Trap}
A cryogenic Penning trap like the one presently discussed can simultaneously confine atomic ions of any charge state of a given element by a combination of a homogeneous static magnetic field and a static electric quadrupole field for periods of days and longer \cite{19}. In general, a Penning trap is similar to an electron-beam ion trap when the electron beam is absent. In turn, a Penning trap can act as an EBIT when an electron beam is added to it, particularly at low electron currents that neither disturb the confining fields nor require an electron collector or electron beam compression by magnetic field gradients. 

In a Penning trap, the confining fields force each individual ion on a bounded trajectory that consists of an oscillation parallel to the magnetic field (the `axial' direction), and of a radial motion perpendicular to it \cite{19}. The frequency $\omega_z$ of the axial oscillation is given by
\begin{equation}
\label{wz}
\omega_z^2=\frac{qU}{md^2} 
\end{equation} 
where $q$ is the charge of the ion, $m$ is its mass, $U$ is the electrostatic potential of the trap, and $d$ its size parameter \cite{19}. Hence, for a given element, each charge state can be uniquely identified by its axial oscillation frequency. The distribution of axial oscillation frequencies can be measured non-destructively, such that the charge-state distribution can be determined during confinement. The relation (\ref{wz}) between charge state and axial oscillation frequency also allows for a selection of a desired charge state by resonant removal of all other charge states by a frequency-selective method called `SWIFT' \cite{20,21}. 

\subsection{Ion Selection by SWIFT}
Any combination of charge-to-mass ratios can be selected to remain confined in the trap by resonant dipole excitation of all other species to the point where they are lost from confinement. In the SWIFT method (Stored Waveform Inverse Fourier Transform), a voltage transient signal is applied to the trap, the Fourier transform of which contains the oscillation frequencies of all undesired species \cite{20,21}. This leads to simultaneous resonant dipole excitation of those unwanted ions. Additionally, the trap depth $U$ may be lowered upon excitation such that the excited ions leave the trap more easily. SWIFT can be applied to either the radial or axial motion of the ions, usually to single out one species of interest to remain in the trap. After each SWIFT cycle, a mass-to-charge spectrum can be taken by non-destructive ion detection to optimise the routine. Presently, SWIFT is performed such that the unwanted ions leave the trap axially along the magnetic field lines and are lost from confinement.   

\subsection{Non-destructive Ion Detection}
The ion content of the trap can be analysed by the signals that the ions' axial motions induce in a dedicated pick-up electrode and a connected resonant circuit (RLC circuit) fixed to a specific frequency
\begin{equation}
\omega_R^2=\frac{1}{LC},    
\end{equation}
by the given inductance $L$ and capacitance $C$ of the circuit. The axial ion motions induce image currents in the pick-up electrode \cite{shock} and thus a voltage across the detection circuit that has a sharp maximum when the actual ion oscillation frequency equals the circuit's resonance frequency, i.e. when we have $\omega_z=\omega_R$. When the trap potential $U$ is ramped across a certain voltage range, the axial oscillation frequencies of all confined ion species are subsequently brought into resonance, hence creating a spectrum of charge-to-mass ratios present inside the trap \cite{19}.

\section{Setup and Procedure}
\subsection{Overview}
\begin{figure}[hhh]
\centering
\includegraphics[width=\columnwidth]{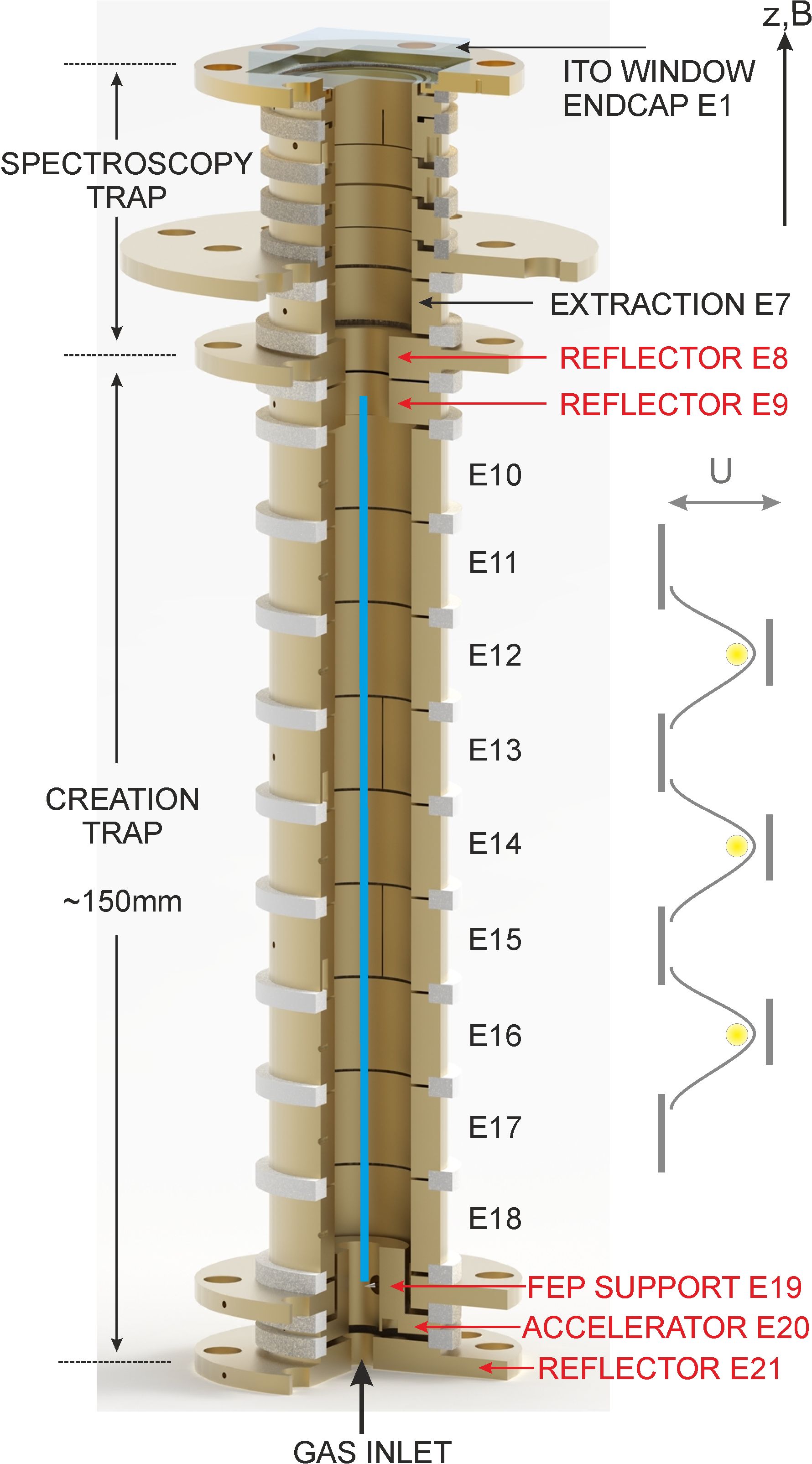}
\caption{\label{one}Sectional view of the Penning trap arrangement: spectroscopy trap on the top, creation trap on the bottom. The central path of the electron beam (light blue) has been added for illustration, so has the triple-well potential for storage of the highly charged ions on the right hand side.}
\end{figure}

The ARTEMIS trap setup, as depicted in figure \ref{one}, consists of a Penning trap for spectroscopy of highly charged ions, and of an adjacent triple-well creation trap in which the highly charged ions are created in close similarity to an EBIT. It is located in the homogeneous field region of a 7\,T superconducting magnet and is cooled to liquid-helium temperature by a commercial cryo-cooler. 
The trap electrode stack consists of 21 electrodes overall, and has an Indium-Tin-Oxide (ITO) coated conducting window at its upper end (`E1'). This window represents an electrically compensated yet optically open endcap, thus forming a half-open trap structure with both a highly harmonic trap potential and favourable light collection properties for spectroscopy \cite{lind,marco}. The mechanically compensated triple-well creation trap features three consecutive harmonic traps for the ions during creation. 
The ambient temperature of about 4\,K ensures efficient cryo-pumping of residual gases in the trap arrangement. From a non-destructive measurement of the ion signal as a function of time, a charge-state lifetime (half-life) for Ar$^{13+}$ of 22 days has been extracted, which indicates a residual gas pressure on the scale of a few times $10^{-16}$\,hPa \cite{ebra}.

\subsection{Electron Source and EBIT Operation}
The EBIT functionality of the creation trap is constituted by the five electrodes indicated in red (figure \ref{one}), namely the FEP support `E19' and the accelerator `E20' which together define the electron beam current and -energy, and the reflector electrodes at either end (`E8/9' and `E21') that are used to reflect the electron beam up and down, such that it traverses the three trap wells located at E12, E14 and E16 multiple times. 

Field emission and electron acceleration are achieved by setting electrode E19 with the FEP to a high negative potential with respect to ground (of the order of -1\,kV to -2\,kV), and the acceleration electrode E20 to a high positive potential. The resulting electron beam is axially contained in the creation trap by setting the reflection electrodes E8/9 and E21 to a potential that is typically about 0.4\,kV more negative than the FEP voltage. The voltages of electrodes E11 to E17 are chosen such that the wells at E12, E14 and E16 have a depth of around 0.25\,kV to contain the ions during charge breeding. The difference of the applied voltages to the trap electrodes is about twice the well depth due to their cylindrical geometry \cite{19}. The FEP voltage supply is able to measure the electron current emitted from the tip to within a few percent.

\begin{figure}[hhh]
\centering
\includegraphics[width=\columnwidth]{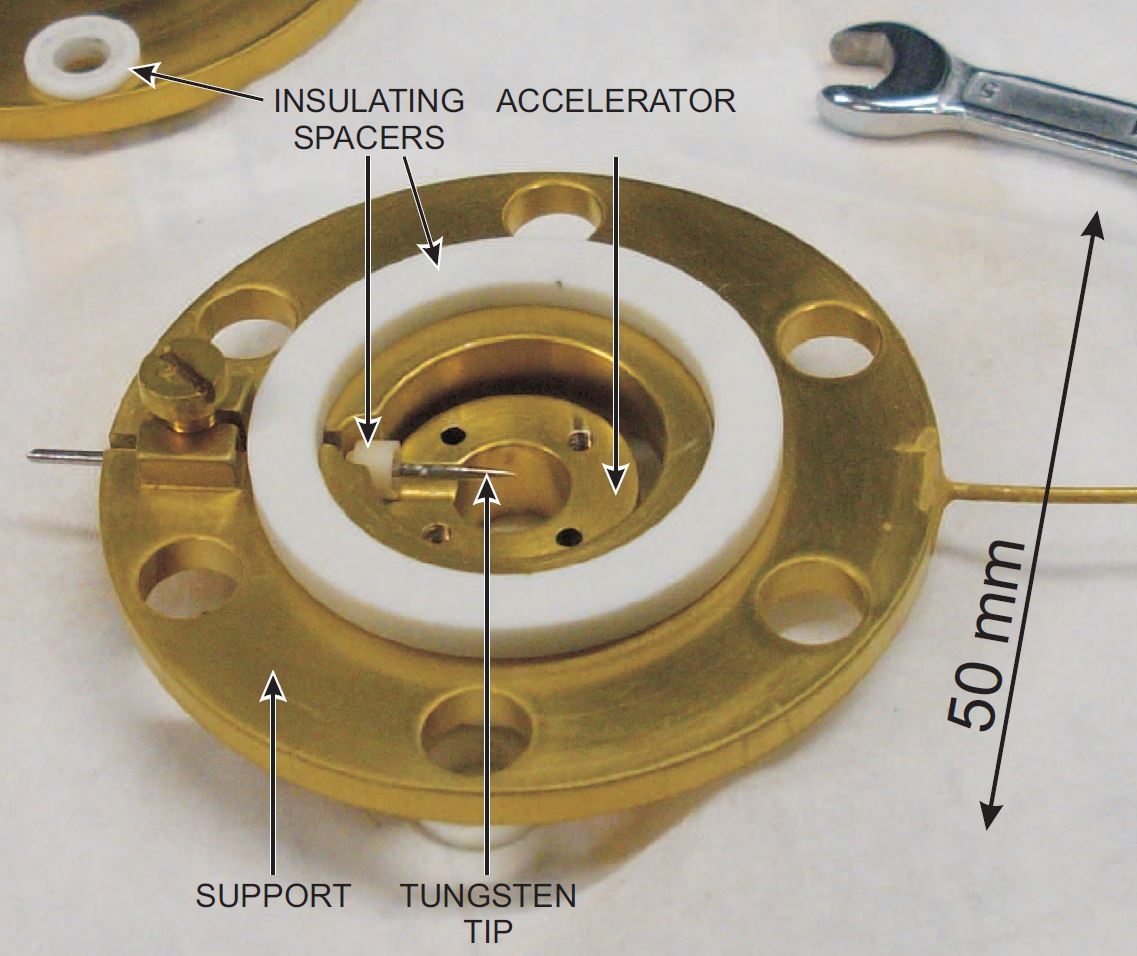} 
\caption{\label{two}Image of the field-emission electron source: support electrode E19 with the needle radially pointing to the central trap axis, and accelerator electrode E20 insulated by a ceramic spacer \cite{dvl}.}
\end{figure}

\begin{figure}[hhh]
\centering
\includegraphics[width=0.7\columnwidth]{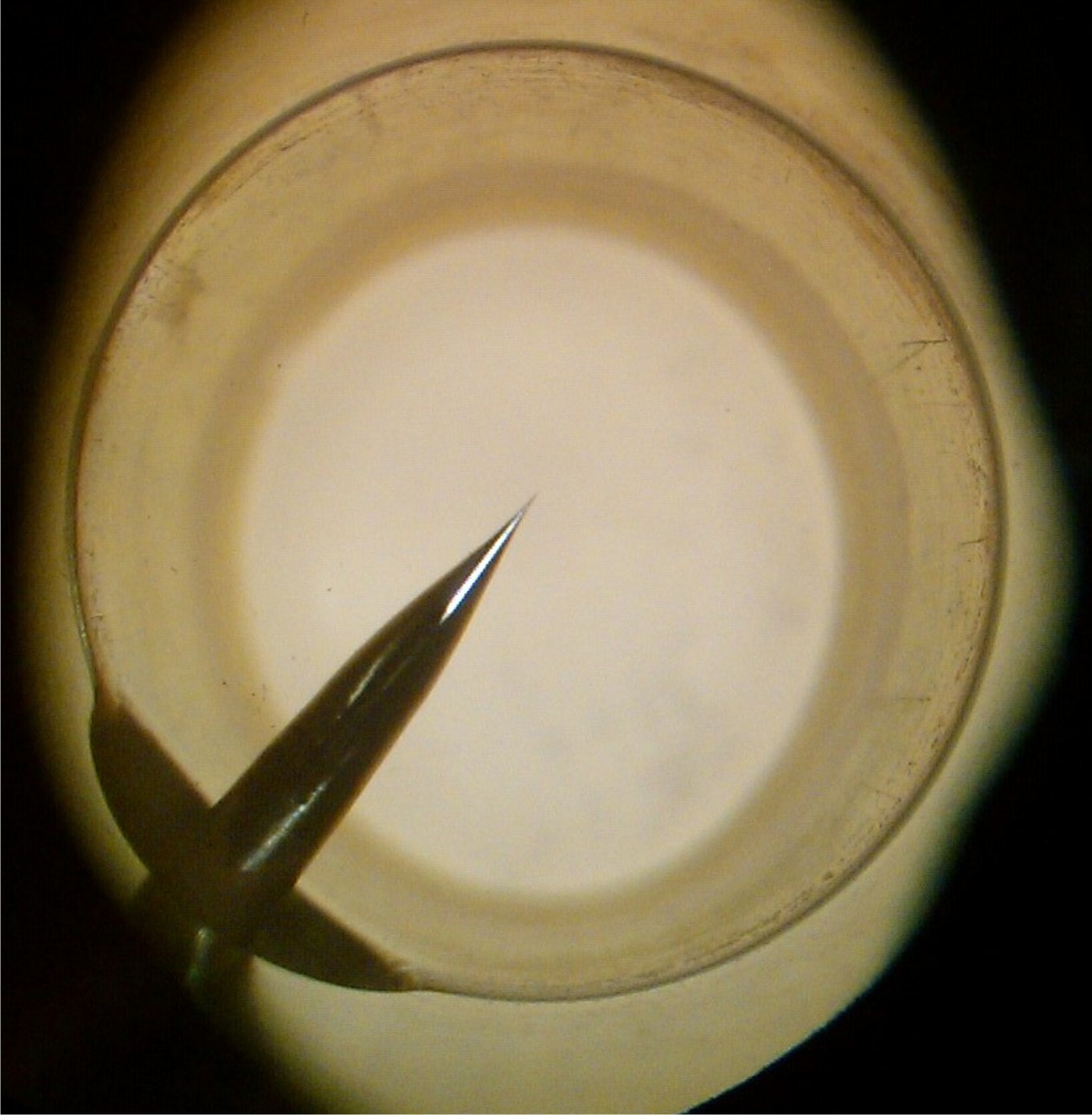} 
\caption{\label{three}Image of the FEP through a light microscope directly from above along the central trap axis. In the background, the accelerator electrode and its central opening for injection of gas or ions from below the trap is visible \cite{dvl}.}
\end{figure}

\begin{figure}[hhh]
\centering
\includegraphics[width=\columnwidth]{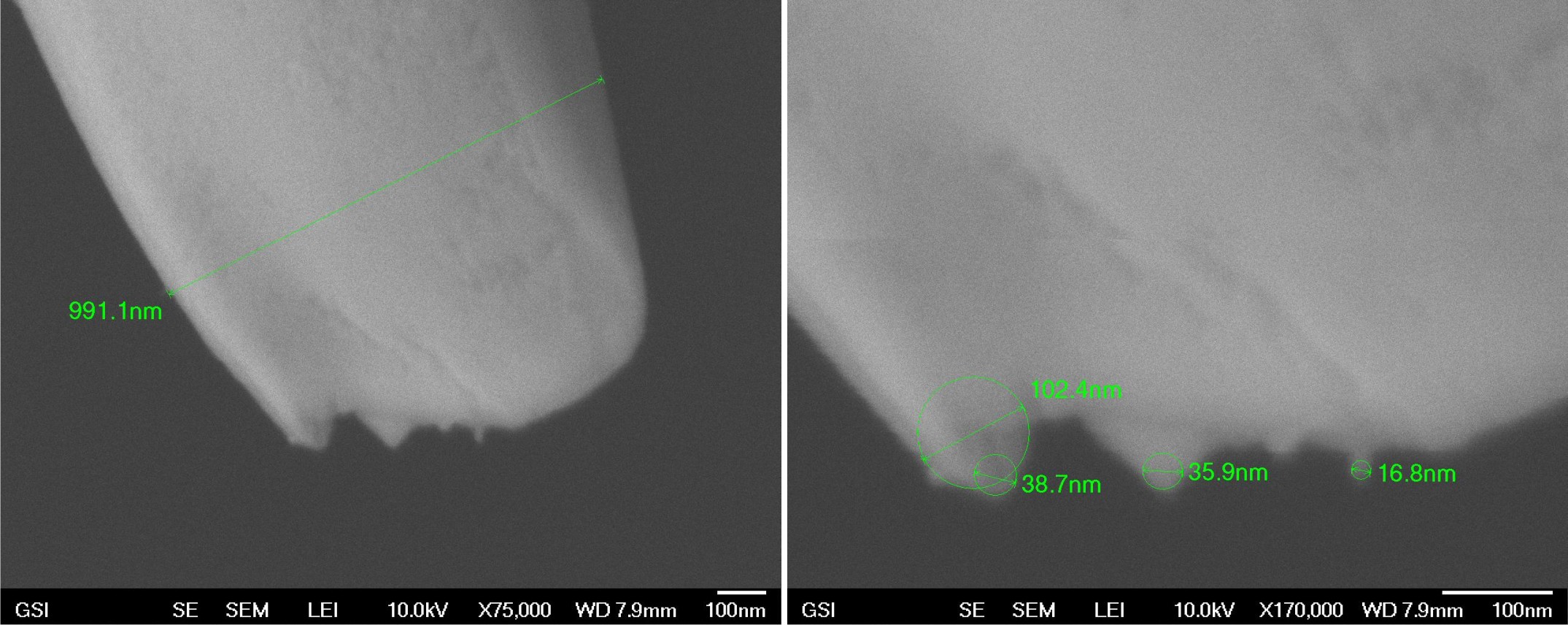} 
\caption{\label{four}Images at different magnifications of the tip under a scanning electron microscope (SEM), with radii of curvature of significant emission points indicated, as measured with the SEM \cite{dvl}.}
\end{figure}

The FEP is located at the sharp end of a needle that radially penetrates the support electrode E19 such that the tip ends on the central trap axis which is also the central magnetic field axis. Figure \ref{two} shows the needle supported by electrode E19 and the accelerator electrode E20 that surrounds the FEP and is insulated from it by a ceramic spacer \cite{dvl}. The electrons emitted from the tip are guided along the central axis by the magnetic field and are axially confined by the voltages applied to the reflector electrodes.  
Figure \ref{three} gives a closer look at the FEP through a light microscope directly from above along the central trap axis \cite{dvl}. In the background, the accelerator electrode and its central opening for injection of gas or ions from below the trap is visible.
The needle with the FEP is produced from a tungsten wire in a specific etching process that results in tips with radii of curvature of 100\,nm and less \cite{dvl}. Figure \ref{four} shows two images at different magnifications of the tip under a scanning electron microscope (SEM), with radii of curvature of significant emission points indicated, as measured with the SEM \cite{dvl}.

\begin{figure}
\centering
\includegraphics[width=0.8\columnwidth]{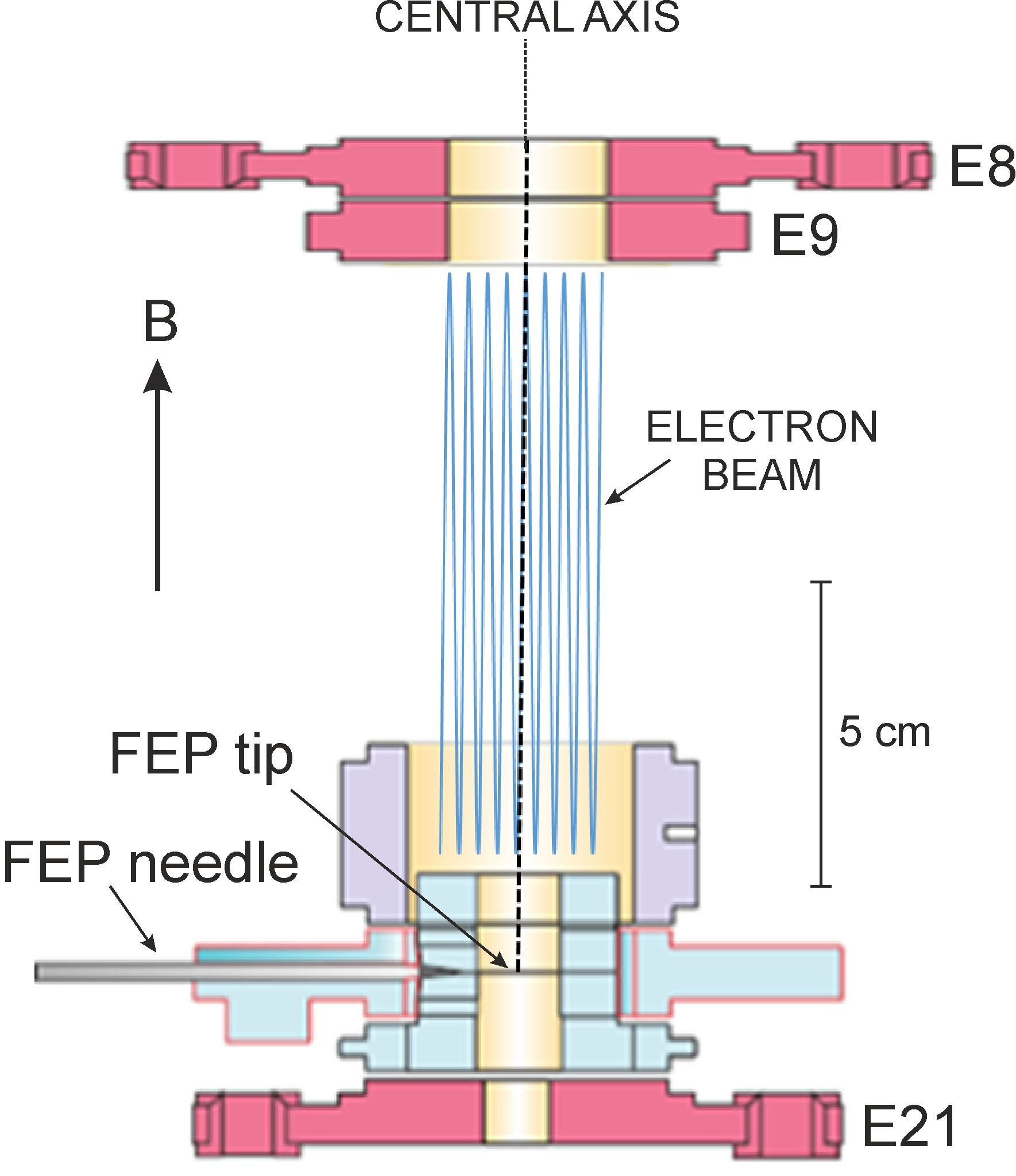} 
\caption{\label{refl}Sketch of the creation part of the setup relevant for electron beam handling, with the electron beam emitted from the FEP along the central trap axis being reflected up and down before radially being lost (not to scale).}
\end{figure}

Figure \ref{refl} illustrates the part of the trap setup relevant for handling of the electron beam, and gives an impression of the electron beam being emitted from the FEP and then reflected up and down multiple times before being lost radially. We will discuss this situation in more detail below.

Note that the existing system can be used to charge-breed gas, other atoms that are present in the trap arrangement (such as atoms dissociated from the tip or sputtered from electrodes hit by the electron beam), and also to further charge-breed ions that have been produced externally and have been dynamically captured into the creation trap. This may not be an immediate advantage when looking at powerful external sources such as the offline ion sources located at HITRAP, or the HITRAP facility itself \cite{andel}, but makes the use of dedicated external ion sources for rare species possible that are otherwise not readily available.

\subsection{Cryogenic Gas Source}
\begin{figure}[hhh]
\centering
\includegraphics[width=0.8\columnwidth]{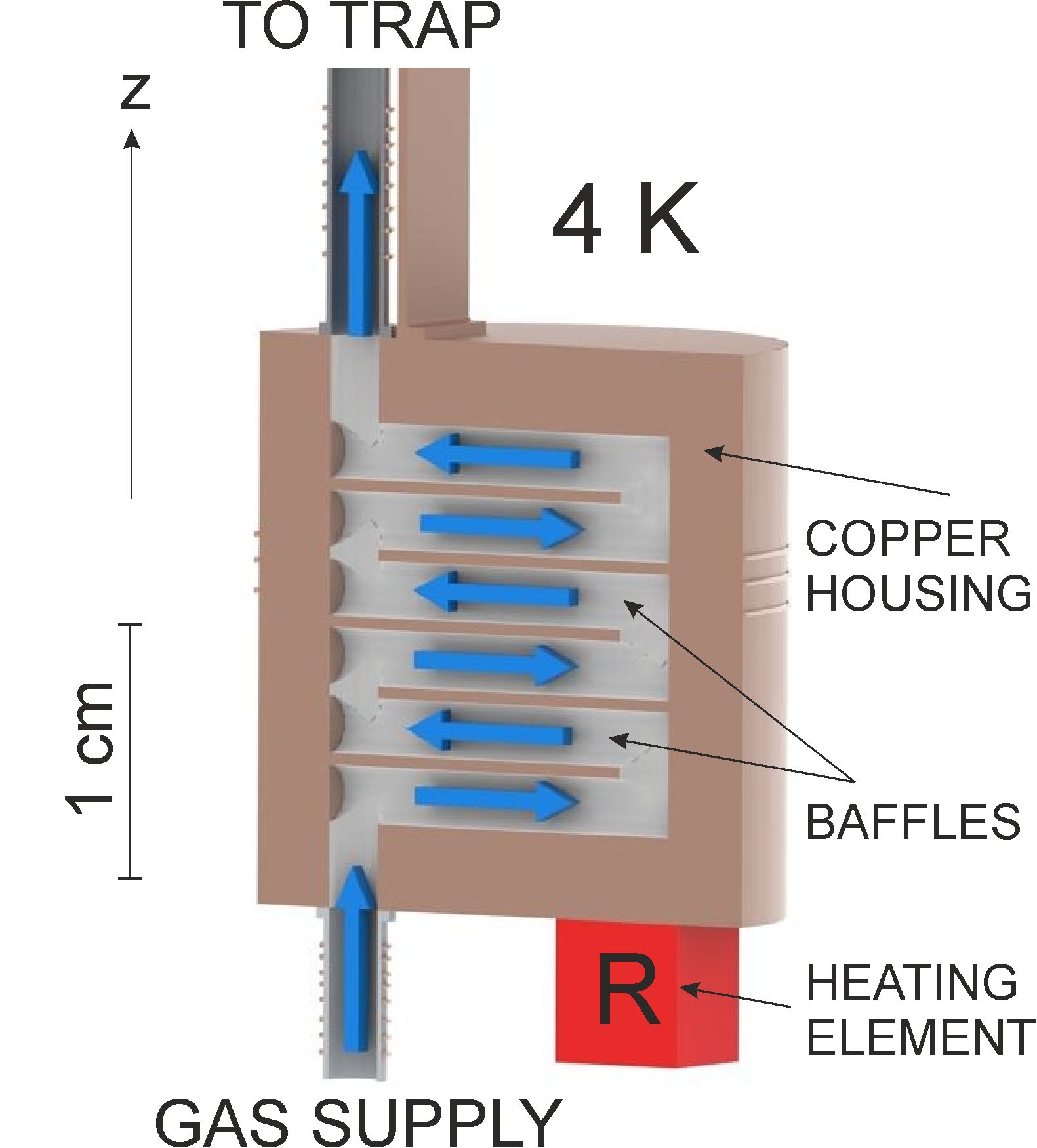} 
\caption{\label{six}Schematic of the cold gas source used to supply small quantities of gas for ionisation by heating it resistively above the temperature required for efficient cryo-pumping of the gas, usually around 30\,K to 40\,K.}
\end{figure}

The gas from which the highly charged argon ions are created comes from a dedicated cold gas source below the trap, which can be heated by a current through a resistor to a temperature typically 30\,K above the ambient temperature of about 4\,K, hence releasing some of the gas frozen inside it. A sectional schematic of this source is depicted in figure \ref{six}.
The cold gas source is prepared prior to ion creation by filling it with the desired gas while it is kept at liquid-helium temperature, such that the gas freezes inside the baffle structure. The baffles form a chicane that efficiently blocks further gas flow from the supply and enables a vacuum better than 10$^{-15}$\,hPa in the trap chamber. When gas is needed, the source is briefly heated by a current through a resistor to a temperature of around 30\,K to 40\,K, such that the gas frozen inside the chicane structure is released into the trap chamber. The value of the current and the corresponding heating are negligible in the given system and neither affect the superconducting magnet nor the cryogenic conditions of the trap, particularly at the typical small duty cycle of the order of a few seconds per day.

\subsection{Ion Detection, Selection, Cooling and Extraction}
Upon ion creation, the ions from the three potential wells located at electrodes E12, E14 and E16 are combined in the middle potential well at E14 by slow switching of the voltages of electrodes E11 to E17 from the configuration shown in figure \ref{one} (right) to a single well located at E14.

Non-destructive ion detection in the creation trap is achieved via a radio-frequency resonator that uses electrode E13 as a pick-up for the axial ion motion and produces a voltage signal that is amplified and read out. At liquid-helium temperature, the resonator has a resonance frequency of $\omega_R=2\pi \times 705.7$\,kHz and a quality factor of $Q=375$. All ion species are brought into resonance subsequently by scanning the creation trap potential $U$, yielding a charge-to-mass spectrum of the trap content, e.g. the one shown in figure \ref{nine}.

Selection of a specific ion species by its charge-to-mass ratio is possible by resonant ejection of all undesired ion species via the SWIFT method. 
The remaining ions are cooled resistively by thermalisation with the resonance circuit at liquid-helium temperature \cite{ebra}. 
Upon selection and cooling, by slow switching of the voltages around the extraction electrode E7, the ions are transferred to the spectroscopy trap for further cooling and / or measurements.
Depending on the details of the production, i.e. amount of gas, breeding current and time, the number density of produced and cooled ions is typically between $10^3$\,cm$^{-3}$ and $10^6$\,cm$^{-3}$, as has been determined from the observed space-charge shift of the axial frequency distribution in a separate set of measurements \cite{ebra}.

\section{Measurements}
\subsection{Electron Current from the FEP}
\begin{figure}[hhh]
\centering
\includegraphics[width=0.9\columnwidth]{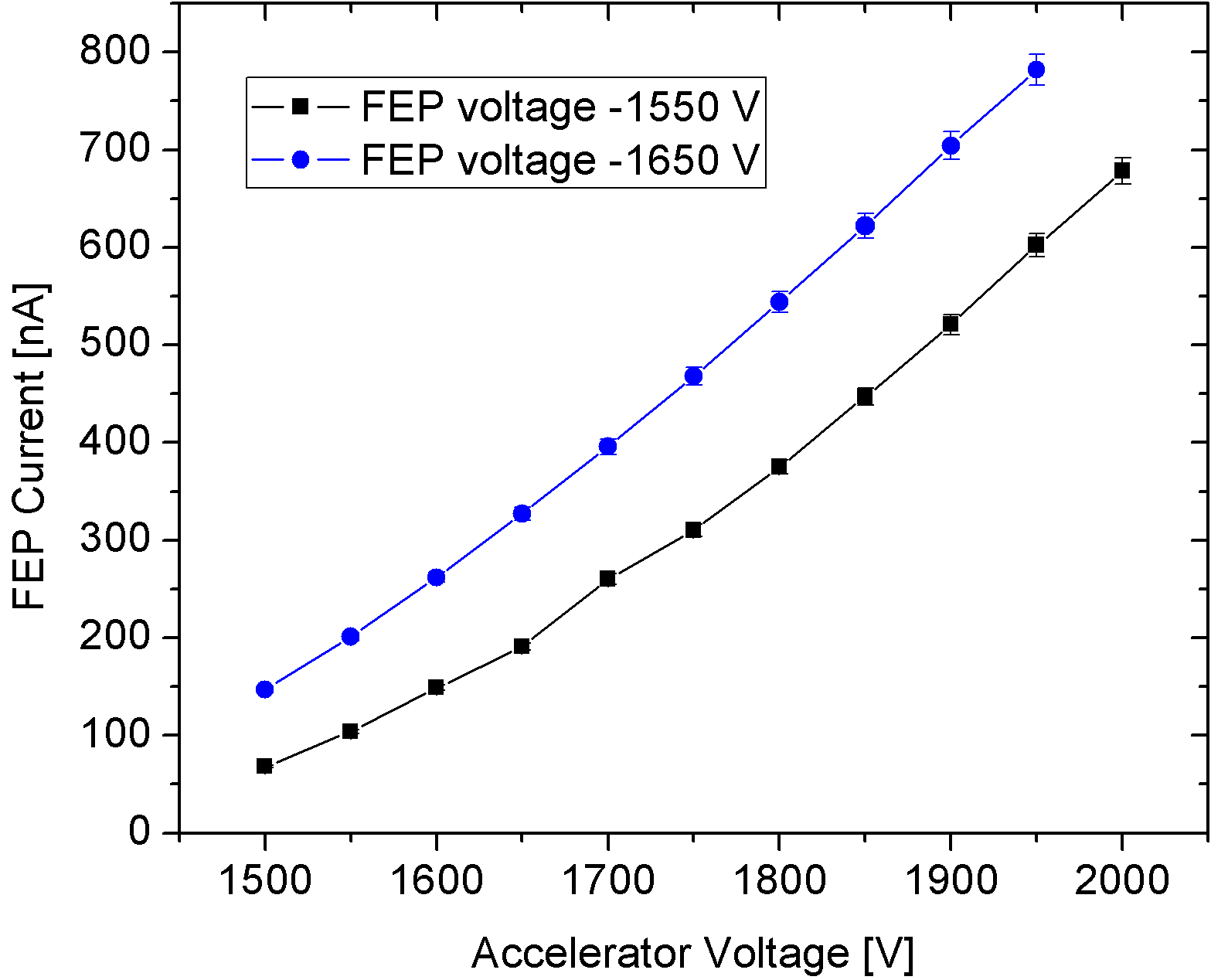} 
\caption{\label{seven}Measured electron current emitted from the FEP as a function of the voltage at the accelerator electrode E20 for two different values of the voltage at the FEP support electrode E19.}
\end{figure}

To characterise the field emission from the FEP, the emitted electron current has been measured as a function of the voltage at the accelerator electrode E20 for two different values of the voltage at the FEP support electrode E19. This is shown in figure \ref{seven}.
As expected, the emitted current increases with increasing FEP voltage with respect to ground and with increasing voltage with respect to the acceleration electrode. In offline tests, FEP tips have shown signs of rapid degradation for currents on the $\mu$A scale, hence we operate the electron source at voltage combinations that restrict the electron current to a few hundreds of nA. 

The duration for which the electron beam is switched on is defined by the desired breeding time. Longer breeding times lead to production of higher charge states but causes stronger heating of the ion cloud. The creation parameters of FEP voltage, accelerator voltage, breeding time and resulting FEP current were varied for optimised ion production, keeping the current low to avoid overly fast degradation of the tip.

Overall, the production performance of the setup is limited mainly by this restriction of the electron current and by the breakdown voltages of the electrode arrangement and cabling which at present give an upper bound of the electron beam energy at roughly 2\,keV. If necessary however, both these limitations could be overcome by use of different kinds of field emitters and by changing design geometries and materials to allow higher voltages to be applied.

\subsection{Gas from the Cold Source}
To characterise the cold gas source, a measurement of the resulting gas pressure inside the vacuum chamber as a function of the heating temperature has been performed prior to installation of the trap. 
\begin{figure}[hhh]
\centering
\includegraphics[width=\columnwidth]{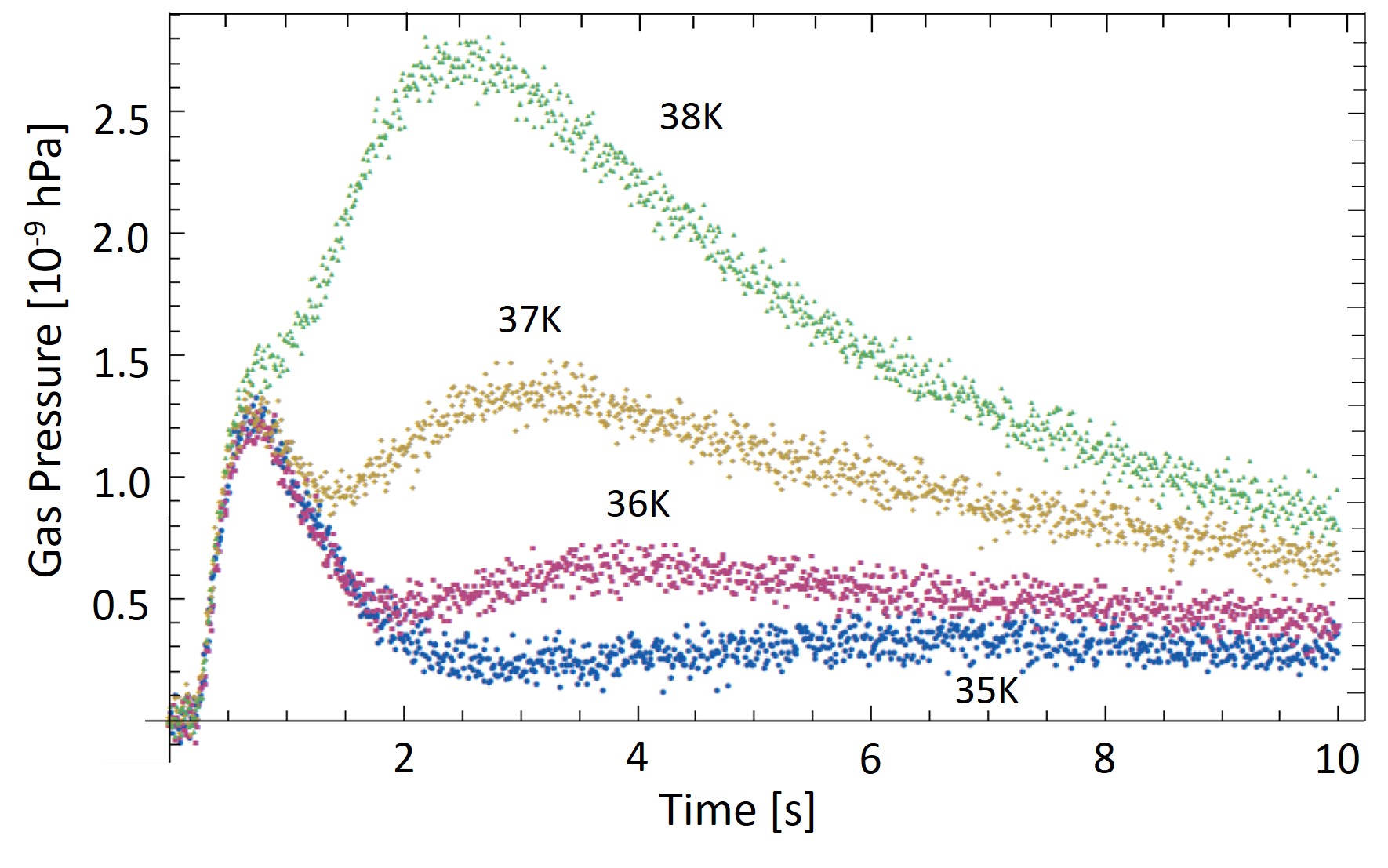} 
\caption{\label{eight}Gas pressure inside the vacuum chamber as a function of time for four different cold-source heating temperatures \cite{dvl}.}
\end{figure}
 Figure \ref{eight} shows the gas pressure in the trap chamber as a function of time for four different values of the initial heating temperature in a pulse of 1\,s duration \cite{dvl}. No gas is detectable for temperatures below 34\,K. Above this temperature, the behaviour is non-linear and the amount of released gas depends critically on the heating, as one expects from general thermodynamic theory \cite{therm}. Upon the initial release during the heating pulse, the gas is efficiently cryo-pumped by the surrounding surfaces at liquid-helium temperature on the time scale of seconds, which is sufficient for ion creation.  

\subsection{Production of Highly Charged Ions: Low Current}
\begin{figure}[hhh]
\centering
\includegraphics[width=\columnwidth]{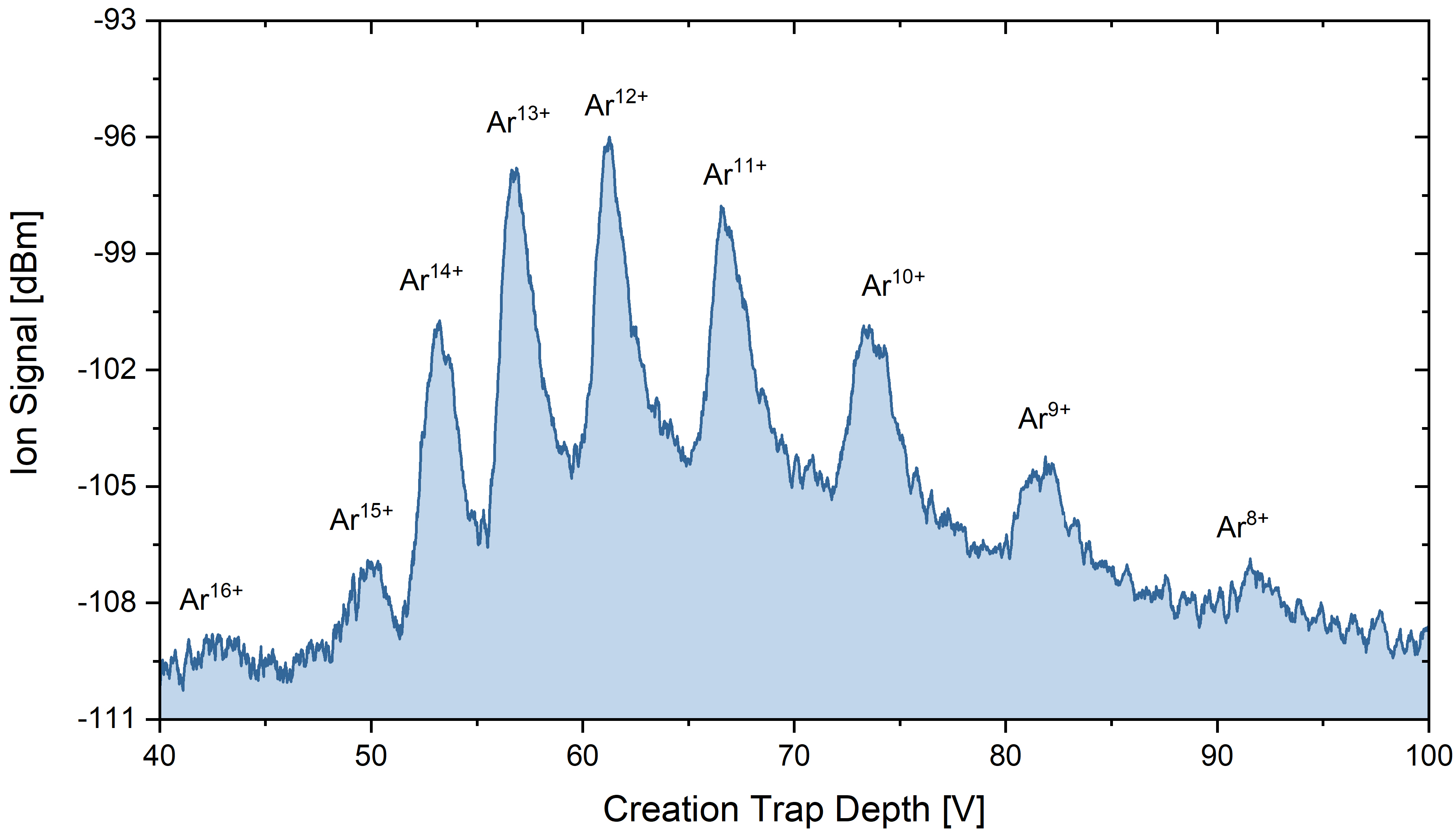} 
\caption{\label{nine}Measured charge-state spectrum of highly charged argon after electron-beam ionisation for $t=1$\,s at an electron beam energy of $E=950$\,eV and a measured electron current from the FEP of $I=170$\,nA.}
\end{figure}

Figure \ref{nine} shows the measured charge-state distribution of highly charged argon ions after electron-beam ionisation for $t=1$\,s at an electron beam energy of $E=950$\,eV and a measured electron current from the FEP of $I=170$\,nA.
The main charge states present in this example are Ar$^{8+}$ to Ar$^{16+}$, peaking around Ar$^{12+}$ and Ar$^{13+}$. This is in agreement with the corresponding ionisation potentials which are all accessible at the given electron beam energy of $E=950$\,eV, namely 143\,eV for Ar$^{8+}$ to 918\,eV for Ar$^{16+}$, whereas the next-higher charge state is inaccessible at 4121\,eV for Ar$^{17+}$ \cite{crc}. Note, that in this measurement, a resonator at a slightly different resonance frequency $\omega_R=2\pi \times 737$\,kHz was used, making the ion peaks appear at slightly different voltages than in the following cases. 

\subsection{Production of Highly Charged Ions: High Current}
\begin{figure}[hhh]
\centering
\includegraphics[width=\columnwidth]{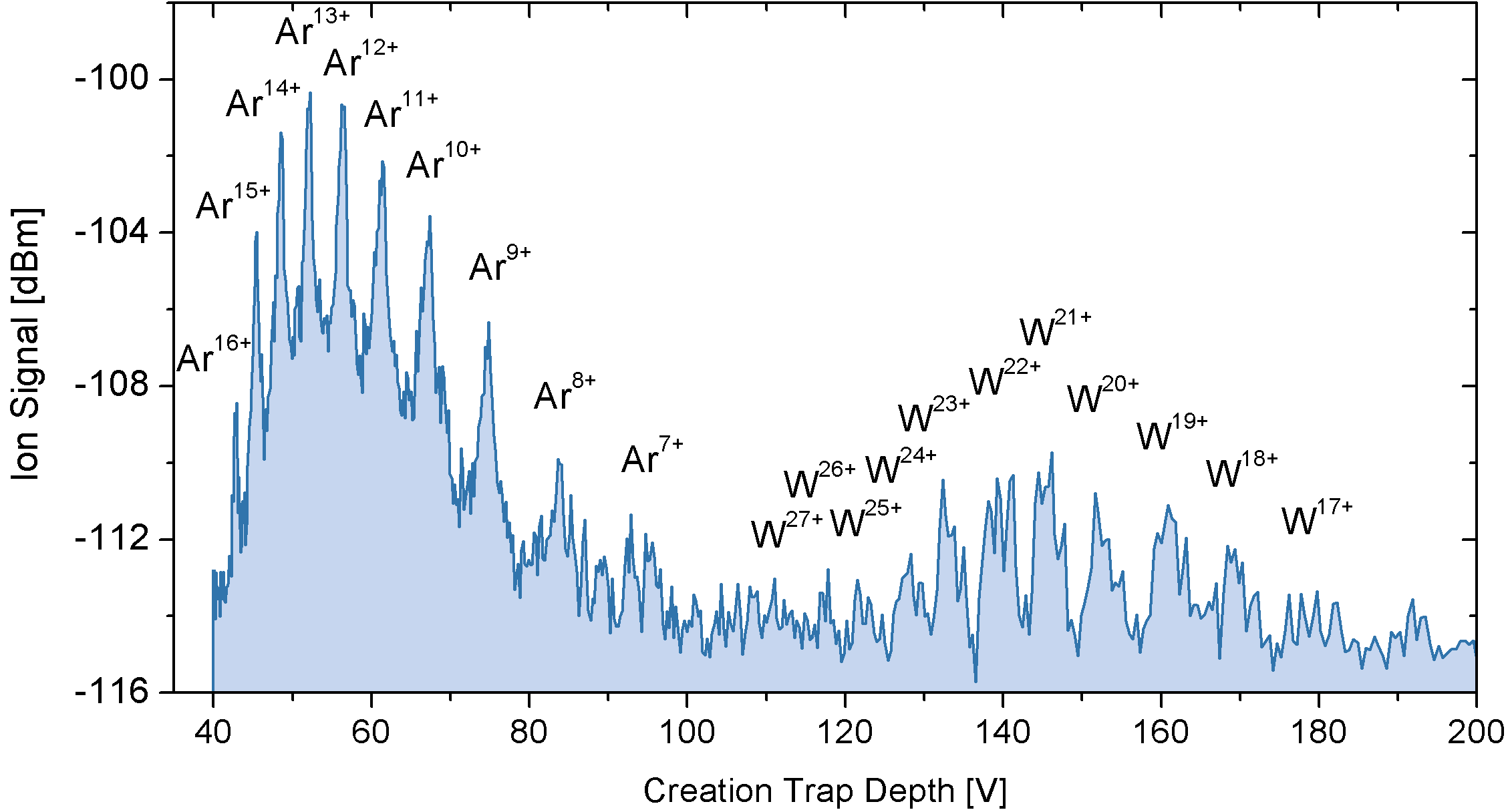} 
\caption{\label{eleven}Charge-to-mass spectrum that includes tungsten ions that have been produced from atoms dissociated from the FEP needle.}
\end{figure}

Figure \ref{eleven} shows a similar charge-to-mass spectrum of argon ions around Ar$^{7+}$ to Ar$^{16+}$ that additionally includes tungsten ions that have been produced from atoms dissociated from the FEP needle at a higher emission current of $I=275$\,nA.
Tungsten ions up to W$^{27+}$ are present, which again is in agreement with the accessible ionisation potentials at the given electron beam energy of $E=950$\,eV, namely up to 881\,eV for W$^{27+}$, whereas the next-higher charge state is inaccessible at 1132\,eV for W$^{28+}$ \cite{crc}.

\subsection{Electron Beam Reflection}
The electron beam reflection used in the current setup is a significant deviation from standard EBIT operation. Hence, we want to have a closer look at the situation. A qualitative illustration of the electron beam arrangement has been given in figure \ref{refl}.

\begin{figure}[hhh]
\centering
\includegraphics[width=\columnwidth]{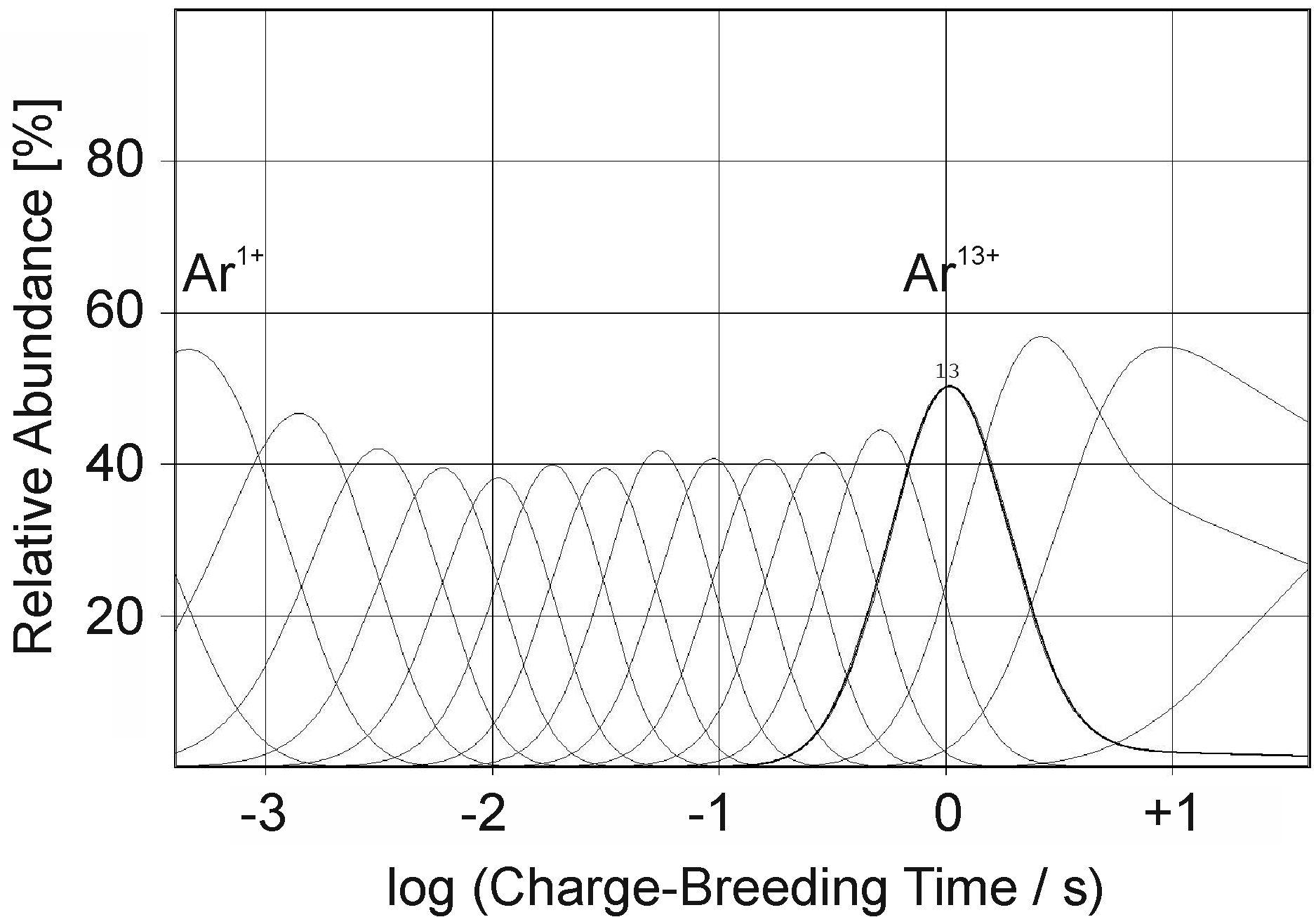} 
\caption{\label{ten}Distribution of relative ion yields as a function of time during electron-beam ionisation at an electron beam energy of $E=950$\,eV and a current density of 2.5\,A/cm$^2$ according to the CBSIM software.}
\end{figure}

A simulation by use of the CBSIM software \cite{17,18} for the conditions of the production shown in figure \ref{nine} has been performed, see figure \ref{ten}. To roughly reproduce the experimentally observed charge states of argon, i.e. a maximum yield around Ar$^{13+}$ at a time of about $t=1$\,s, an electron current density $j$ of about 2.5\,A/cm$^2$ is required. We take this as an indication that the actual electron beam in our setup, being reflected up and down multiple times during charge breeding, leads to the same ion production as a single-pass electron beam in a conventional EBIT of that current density.   

Looking down on the ion creation arrangement along the central trap axis, the light distribution emitted from its top has a diameter of about 1\,mm and is understood to be mainly Bremsstrahlung from the electrons being reflected repeatedly during charge breeding \cite{lind}. From the measured electron current of $I=170$\,nA and the current density of $j=2.5$\,A/cm$^2$, one would obtain a cross-sectional area $A=I/j$ of a single-pass electron beam of roughly $10^{-7}$\,cm$^2$. This area is smaller than the measured light distribution by a factor of about $10^6$, which indicates that our electron beam is reflected up and down a correspondingly large number of times while being radially expanded due to space charge effects. 

We can obtain a rough number of electron beam reflections also from assuming that the reflected electron beam fills up the trap to the point where radial loss occurs at the rate of electron production, i.e. the trap is filled to the Brillouin limit at which space charge overcomes the magnetic confinement of the trap \cite{19}. This electron number density is given by
\begin{equation}
    n=\frac{\epsilon_0 B^2}{2m}
\end{equation}
where $\epsilon_0$ is the permittivity of free space and $m$ is the electron mass. At the present field of $B=7$\,T, we have $n \approx 10^9$/cm$^3$. The total charge in the trap is then given by $Q=enV$ which for our trap with $V \approx 10$\,cm$^3$ is about $Q \approx 10^{-9}$\,C. The FEP current of roughly 100\,nA takes about 10\,ms to fill the trap to that limit, during which time electrons at 1\,keV energy travel $10^5$\,m. At the length of our trap of about 10\,cm this means a number of reflections of about $10^6$ which agrees with the above assumption. 

During charge-breeding, the currents on the FEP and accelerator electrodes are observed to be equivalent. This is expected as the accelerator has the smallest inner diameter and an attractive potential, and thus acts similarly to the collector electrode in a typical EBIT operating in reflex mode. The average lifetime at the Brillouin limit of an electron in the trap can be considered from emission at the FEP until impact with the accelerator electrode. From the current and the total number of trapped electrons, it is determined to be about 10\,ms. During this time, many mutual Coulomb scatterings lead to a random walk of electrons from the center radial position, such that longer-lived electrons occupy larger radial positions on average. After reaching a steady state, the outermost radial electrons may be ejected by the space charge of the inner electrons. In this way, newly emitted electrons are vastly more likely to replace an older electron in the plasma than to be directly ejected. We assume the initial acceleration after emission to dominate the electron kinetic energy, yet due to electron-electron interaction occurring in the reflex mode, the beam may not be fully mono-energetic.

\subsection{Charge-State Selection}
Selection of a single charge state is performed by resonant ejection of all other species upon creation. The corresponding SWIFT excitation signal to eject the unwanted ions axially is irradiated via electrode E13 which is the axial neighbour of E14 where the ions are located.
Since in the present setup the SWIFT excitation voltage is limited to 10\,V, the excitation signal is applied repeatedly until the unwanted ions are removed. During the irradiation, electrode E14 is kept at -250\,V and the adjacent electrodes are at +10\,V, giving a trap depth of 260\,V. This method of selection can be combined with a short-term lowering of the trap depth, thus allowing excited ions to leave the trap more easily while maintaining the desired ions.
\begin{figure}[hhh]
\centering
\includegraphics[width=\columnwidth]{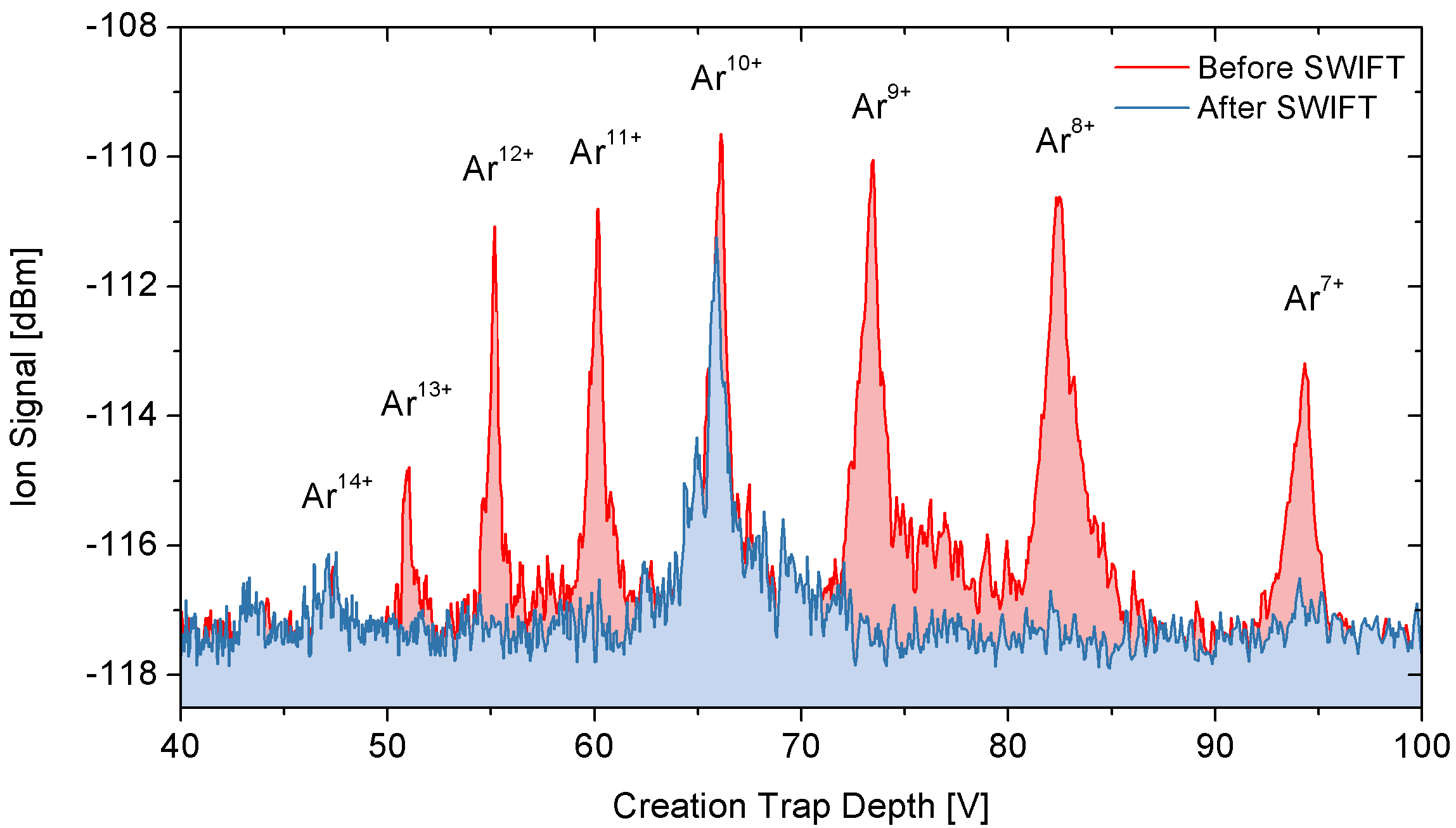} 
\caption{\label{twelve}Charge-to-mass spectrum of argon ions before and after selection of Ar$^{10+}$ (blue) by SWIFT removal of all other species (red) from the trap.}
\end{figure}
Figure \ref{twelve} shows a charge-to-mass spectrum of argon ions before and after selection of Ar$^{10+}$ by SWIFT removal of all other species from the trap. 
While the undesired charge states (red) are removed completely, about 95\,\% of the selected Ar$^{10+}$ (blue) are still present, proving that spurious excitation of the selected species is small, and that SWIFT is efficient for charge-state selection in a cooled cloud of ions.

\section{Summary and Conclusion}
We have described the setup, applied methods, and operation of a device for the production, selection and storage of highly charged ions inside a cryogenic Penning trap. Ion production takes place by charge-breeding of atoms via impact ionisation with electrons from a cryogenic field emitter. We have shown the efficient production of highly charged ions up to ionisation potentials of the order of one keV on the time scale of seconds from gas injected by a dedicated cold gas source. The efficiency mainly goes back to the reflex mode of operation, in which the electron beam is re-used a large number of times. Upon creation, the charge-state distribution is analysed non-destructively, and desired charge states are selected for further study by resonant removal of unwanted ions and by adiabatic transport of the ions of interest to an adjacent Penning trap for spectroscopic studies.   

The methods and device described here may be used beyond the specific application above since it allows to create and select highly charged ions across a wide range of species, charge states and total ion numbers, and to create them at low kinetic energy when compared to typical situations with dynamic (in-flight) capture \cite{schn}. Also, it allows for operation in cryogenic environments that are typically required for precision trap experiments as they allow the use of superconducting equipment, low-noise electronics for detection and manipulation of ions, and particularly extreme vacua that allow for long ion storage times of days and more as they are required for many precision studies with highly charged ions \cite{1,8,p1,p2,p3}.

\section*{Acknowledgments}
We acknowledge financial support by the Federal Ministry of Education and Research (BMBF) [Grants 05P21RDFA1 and 05P2021 (ErUM-FSP T05)], by the Helmholtz Forschungsakademie Hessen für FAIR (HFHF), the R\&D cooperation agreement between GSI/FAIR and Heidelberg
University, the European Union’s Horizon 2020 research and innovation programme under the Marie Skłodowska-Curie grant agreement No 721559 `Accelerators Validating Antimatter physics (AVA)', the International Max Planck Research School for Quantum Dynamics (IMPRS-QD), and by the Helmholtz Graduate School for Hadron and Ion Research (HGS-HIRe for FAIR). We further acknowledge the work done by the previous PhD students D. von Lindenfels and M. Wiesel whose efforts paved the way for the presented work.


\section*{References}
\begin{thebibliography}{99}
\bibitem{null} H F Beyer and V P Shevelko, Introduction to the Physics of Highly Charged Ions, CRC Press (2002)
\bibitem{3} T Beier, Physics Reports 339, 79 (2000)
\bibitem{micke} P Micke, T Leopold, S King, E Benkler, L J Spiess, L Schmöger, M Schwarz, J R Crespo Lopez-Urrutia and P O Schmidt, Nature 578, 60 (2020)
\bibitem{gumb} A Gumberidze, Th Stöhlker, D Banas, K Beckert, P Beller, H F Beyer, F Bosch, S Hagmann, C Kozhuharov, D Liesen, et al., Phys. Rev. Lett. 94, 223001 (2005)
\bibitem{dumb} P Beiersdorfer, H Chen, D B Thorn, and E Träbert, Phys. Rev. Lett. 95, 233003 (2005)
\bibitem{1} S Sturm, M Vogel, F Köhler-Langes, W Quint, K Blaum and G Werth, Atoms 5, 4 (2017)
\bibitem{2} M Vogel and W Quint, Ann. Phys. (Berlin) 525, 505 (2013)
\bibitem{2b} W Quint, D L Moskovkhin, V M Shabaev and M Vogel, Phys. Rev. A 78, 032517 (2008)
\bibitem{3b} A V Volotka, D A Glazov, G Plunien, V M Shabaev, Ann. Phys. (Berlin) 525, 636 (2013)
\bibitem{3c} V M Shabaev, A I Bondarev, D A Glazov, M Y Kaygorodov, Y S Kozhedub, I A Maltsev, A V Malyshev, R V Popov, I I Tupitsyn and N A Zubova, Hyp. Int. 239, 60 (2018)
\bibitem{4} M G Kozlov, M S Safronova, J R Crespo Lopez-Urrutia and P O Schmidt, Rev. Mod. Phys. 90, 045005 (2018)
\bibitem{5} J C Berengut, V A Dzuba, V V Flambaum and A Ong, Phys. Rev. Lett. 106, 210802 (2011) 
\bibitem{6} S Sturm, F Köhler, J Zatorski, A Wagner, Z Harman, G Werth, W Quint, C H Keitel and K Blaum, Nature 506, 467 (2014)
\bibitem{19} M Vogel, Particle Confinement in Penning Traps, Springer Series on Atomic, Optical, and Plasma Physics 100, Springer (2018)
\bibitem{7} L Gruber, J P Holder and D Schneider, Physica Scripta 71, 60 (2005)
\bibitem{8} S Sturm, I Arapoglou, A Egl, M Höcker, S Kraemer, T Sailer, B Tu, A Weigel, R Wolf, J Crespo Lopez-Urrutia and K Blaum , Eur. Phys. J. Spec. Top. 227, 1425 (2019)
\bibitem{9} T Murböck, S Schmidt, G Birkl, W Nörtershäuser, R C Thompson and M Vogel, Phys. Rev. A 94, 043410 (2016)
\bibitem{9b} M Kiffer, S Ringleb, N Stallkamp, B Arndt, I Blinov, S Kumar, S Stahl, Th. Stöhlker, and M. Vogel, Rev. Sci. Inst. 90, 113301 (2019)
\bibitem{array} S A Guerrera and A I Akinwande, Nanotechnology 27, 295302 (2016)
\bibitem{10} J Alonso, K Blaum, S Djekic, H J Kluge, W Quint, B Schabinger, S Stahl, J Verdu, M Vogel and G Werth, Rev. Sci. Instrum. 77, 03A901 (2006)
\bibitem{donets} E D Donets, E E Donets and E Syresin, Rev. Sci. Inst. 71, 887 (2000)
\bibitem{art2} D von Lindenfels, M Wiesel, W Quint, D Glazov, V M Shabaev, G Birkl and M Vogel, Phys. Rev. A 87, 023412 (2013)
\bibitem{hitrap} H J Kluge, T Beier, K Blaum, L Dahl, S Eliseev, F Herfurth, B Hofmann, O Kester, S Koszudowski, et al, Advances in Quantum Chemistry 53, 83 (2007)
\bibitem{hit2} F Herfurth, Z Andelkovic, W Barth, W Chen, L A Dahl, S Fedotova, P Gerhard, M Kaiser, O K Kester, H J Kluge, G. Maero et al. Phys. Scr. T166,  014065 (2015)
\bibitem{dvl} D von Lindenfels, PhD thesis, Heidelberg (2015)
\bibitem{ebra} M S Ebrahimi, Z Guo, M Wiesel, G Birkl, W Quint and M Vogel, Phys. Rev. A 98, 023423 (2018)
\bibitem{11} R H Fowler and L Nordheim, Proc. R. Soc. Lond. A 119, 173 (1928)
\bibitem{12} E W Müller, Z. Phys. 106, 541 (1937)
\bibitem{13} M A Levine, R E Marrs, J R Henderson, D A Knapp and M B Schneider, Physica Scripta T22, 157 (1987)
\bibitem{13b} W Lotz, Z. Phys. 206, 205 (1967)
\bibitem{14} B M Penetrante, J N Bardsley, D DeWitt, M Clark and D Schneider, Phys. Rev. A 43, 4861 (1991)
\bibitem{15} Y S Kim and R H Pratt, Phys. Rev. A 27, 2913 (1983)
\bibitem{16} A Müller and E Salzborn, Phys. Lett. 62A, 391 (1977)
\bibitem{17} R Becker and M Kleinod, Rev. Sci. Instrum. 65, 1063 (1994)
\bibitem{18} R Becker, O Kester and Th Stöhlker, Journal of Physics: Conference Series 58, 443 (2007)
\bibitem{20} A G Marshall, C L Hendrickson and G S Jackson, Mass Spectrom. Rev. 17, 1 (1998)
\bibitem{21} S Guan and A G Marshall, Int. J. Mass Spectrom. Ion Proc. 157-158, 5 (1996)
\bibitem{shock} W Shockley, J. Appl. Phys. 9, 635 (1938)
\bibitem{lind} D von Lindenfels, M Vogel, G Birkl, W Quint and M Wiesel, Hyp. Int. 227, 197 (2014)
\bibitem{marco} M Wiesel, G Birkl, M S Ebrahimi, A Martin, W Quint, N Stallkamp and M Vogel, Rev. Sci. Inst. 88, 123101 (2017)
\bibitem{therm} K Wark and D E Richards, Thermodynamics, 6th Ed., McGraw-Hill (1999)
\bibitem{andel} Z Andelkovic, N Kotovskiy, K König, B Maas, T Murböck T, D Neidherr, S Schmidt, J Steinmann, G Vorobjev and F Herfurth, Nucl. Inst. Meth A 795, 109 (2015)
\bibitem{crc} David R Lide, CRC Handbook of Chemistry and Physics, 84th Edition, CRC Press, Boca Raton, (2003)
\bibitem{schn} H. Schnatz et al., Nuclear Instruments and Methods in Physics Research Section A
251, 17 (1986) 
\bibitem{p1} S Sturm et al., Phys. Rev. Lett. 107, 023002 (2011)
\bibitem{p2} A Wagner et al., Phys. Rev. Lett. 110, 033003 (2013)
\bibitem{p3} F Köhler et al., Nat. Comm. 7, 10246 (2016)
\end{thebibliography}
\end{document}